\documentstyle[A4,11pt,psfig]{article}
\newcommand{\beq}{\begin{eqnarray}}
\newcommand{\eeq}{\end{eqnarray}}
\begin{document}
\title{On the group velocity of oscillating neutrino states and the equal velocities assumption}
\author{J.-M. L\'evy \thanks{Laboratoire de Physique Nucl\'eaire et de Hautes Energies,
CNRS - IN2P3 - Universit\'es Paris VI et Paris VII, Paris.   \it Email: jmlevy@in2p3.fr}}
\pagenumbering{arabic}
\sloppy
\maketitle
\begin{abstract}
In a two flavour world, the usual equal $p$, equal $E$, or equal $v$
assumptions used to derive the neutrino oscillation length can be replaced by
a specific definition of the velocity of the oscillating state which, unlike
those assumptions, is compatible with exact production kinematics. 
This definition is further vindicated by the analysis of $\pi_{\mu 2}$ decay at rest.\\
\end{abstract}
\newpage
\section{Introduction}
Contradictory arguments are still exchanged in the discussion of neutrino oscillations in vaccuum 
not to mention more complicated cases. 
People keep arguing about the superiority of their favourite assumption (equal $p$'s, $E$'s, or 
$v$'s \cite{take}) in deriving an oscillation length. Some of the champions of the
last approximation even claim that the usual formula must be corrected by a factor of two 
\cite{deleo} while defenders of the standard result have spent some time showing that the equal $v$'s
is kinematically untenable \cite{ok},\cite{GK} .\\

Although this last assumption is more at variance with 
exact production kinematics than the other two, treating it with care helps 
to uncover a weaker hypothesis which is both sufficient and compatible with kinematics,
at least in a two flavour - two mass neutrino world.\\

Using the naive plane-wave model in this simple case, we shall show that:
\begin{enumerate}
\item The equal $v$ assumption does yield the same 
oscillation length as the other two provided velocity is treated consistently, i.e. the same value is used in
the momentum-energy and in the position-time relations.
\item To derive this oscillation length, the only necessary assumption
 which is compatible with exact kinematics, and is, in some way, contained in the three usual approximations is a 
specific definition of the group velocity of the would-be wave packet.
\item This value of the group velocity must be postulated, but is vindicated by the 
equality of the derived oscillation length with that found by more elaborate methods; 
moreover, it seems difficult to avoid in the case of $\pi \rightarrow \mu \nu$ at rest.

\end{enumerate}
\section{Standard and 'equal velocities' assumptions}
Being confined to a two mass world, we shall use simplified notations from the start
and represent the neutrino born at space-time point $(0,0)$ in some charged current reaction 
involving charged lepton $l = e\ \rm{or} \ \mu $ by: 
\beq |0,0> = |\nu_l> = \cos{\theta}|1> + \sin{\theta}|2>\label{heav} \eeq where $|h> \ h=1,2$ are mass 
eigenstates with eigenvalues $m_h$ and definite energy and momentum $E_h$ and $p_h$.\\
Assuming only the existence of a momentum operator commuting with the
Hamiltonian, (\ref{heav}) can be propagated by use of the space-time 
translation operator ${\cal U}=e^{-i(Ht-\vec{P}\cdot\vec{r})}$ which yields:
\beq {\cal U}|0,0> = |x,t> =  \cos{\theta} e^{-i(E_1t-p_1x)}|1> + \sin{\theta} e^{-i(E_2t-p_2x)}|2>\label{act} \eeq
We have taken the $x$ axis along the direction of motion. The values of $E_h$ and $p_h$ 
depend on the production kinematics; in the case of a $\pi_{l2}$ decay for example, they are
fixed by the masses in the $\pi$ rest frame and from there in the lab, once the $\pi$'s decay angle and velocity are given.
Projection of $|x,t>$ onto $|\nu_{l'}> = -\sin{\theta}|1> + \cos{\theta}|2>$ gives then
\beq A_{l \rightarrow l'}(x,t) = \sin{\theta}\cos{\theta}(e^{-i(E_2t-p_2x)}-e^{-i(E_1t-p_1x)})
 \label{amplit}\eeq
for the amplitude to detect a neutrino of flavor $l'\neq l$ at point $x$ and time $t$ relative to the
production point at $(0,0)$, if that neutrino undergoes a C.C. interaction.
The object of interest is the phase difference $\delta \phi = \delta E t -\delta p x$ 
(with $\delta E = E_1-E_2$ and $\delta p = p_1 - p_2$) which appears 
in the oscillation probability obtained by squaring (\ref{amplit}):
\beq P_{l\rightarrow l'}(x,t) = \sin^2(2\theta) \sin^2( \frac{\delta \phi}{2}) \eeq
To evaluate $\delta \phi$, people postulate equal momenta, or equal energies, (see e.g. \cite{KAY}) or
even equal velocities (e.g. \cite{deleo}) of the two definite mass components. We first show that 
all three assumptions yield the same expression without further approximations: 
\subsection{Equal energies}
\label{eqe}
It is assumed here that $E_1 = E_2$, hence $\delta \phi = -\delta p x .\; $ 
 However the relation: \beq \delta p^2 = \delta E^2 -\delta m^2 \label{core}\eeq
yields in this case: \beq \delta p = -\frac{\delta m^2}{p_1+p_2} = -\frac{\delta m^2}{2 \bar{p}}&
\rm{and}  & \delta \phi = \frac{\delta m^2}{2 \bar{p}} x \nonumber \eeq with an
obvious definition for $\bar{p}$.\\

Requiring then $  \delta \phi = 2\pi $ we find the well known result:
 \beq L_{osc} = \frac{4\pi\bar{p}}{|\delta m^2|}\label{exact} \eeq 
\subsection{Equal momenta} 
\label{eqp}
Here, $\delta \phi = \delta E t$ and ({\ref{core}) yields $\delta E = \frac{\delta m^2}{E_1+E_2}$.
If $v$ is the velocity of the center of the would-be wave packet, one finds:
\beq \delta \phi = \delta E t = \frac{\delta m^2}{E_1+E_2} t = \frac{\delta m^2}{2p}x 
\label{fullphase}\eeq
where the last equality holds exactly provided {\boldmath $v = \frac{2p}{E_1+E_2}$} \\
With this definition for $v$, the same oscillation length as in (\ref{eqe}) is found without 
the traditional ultra-relativistic ingredients used to transform (\ref{fullphase}) 
by setting $t \approx x$ for the neutrino trajectory and 
$E_1+E_2 \approx 2p$ .

\subsection{Equal velocities} 
\label{eqb}
In this case, $\delta \phi$ does not reduce to a single term and it is very important {\bf not} to 
"approximate" $t$ by $x$, for in so doing one would arrive at:
 $$\delta \phi = (\delta E - \delta p)x =  \delta m e^{-\eta}x$$ upon introducing $\eta = \tanh^{-1}(v)$\\

The oscillating pattern would be described by: 
\beq \sin^2(\frac{|\delta m|}{2}e^{-\eta}x)&
\rm{ and\ the\ oscillation\ length:\ } &L'_{osc}=\frac{2\pi e^{\eta}}{|\delta m|}\eeq
However, since $$e^{\eta} = \frac{E_1+p_1}{m_1} = \frac{E_2+p_2}{m_2} \approx \frac{2(E+p)}{m_1+m_2}$$ 
this yields finally
\beq L'_{osc.}=\frac{4(E+p)\pi}{|\delta m^2|} \label{false}\eeq {\it viz.} twice the usual value in the
relativistic regime where $E \approx p$; this is the origin of the claims of some authors.

However, if the hypothesis of equal velocities has any meaning, then the center of the wave packet
moves with {\bf that} velocity, not with velocity 1. Therefore, defining its position by 
$x = v t$ yields:
\beq \delta \phi = \delta E (t-v x) = \delta m \gamma (1/v -v) x = \frac{\delta m}{v \gamma} x \label{truep}\eeq
\footnote{$\gamma = 1/\sqrt{1-v^2}$ as usual}
Now $$\frac{1}{v \gamma} = \frac{m_1}{p_1} = \frac{m_2}{p_2} = \frac{m_1+m_2}{2 \bar{p}}$$
Hence $\delta \phi = \frac{\delta m^2}{2\bar{p}}x$ and the formula found in (\ref{eqe}) results.\\

Clearly, replacing $1/v-v$ by $1-v$ (equivalent to $t \rightarrow x$ in \ref{truep}) cannot be harmless; 
in the situation of \ref{eqp}, $v$ is
only an overall factor, and replacing it by $1$ would induce a relative error on the phase difference which goes to $0$ with 
$1-v$. Not so in the present case where the relative error is $1/(1+v)$ - hence the factor $2$ found above.
Stated differently, the same value for $v$ must be used in configuration and in momentum space.\\
One sees that in \cite{KAY} for example, the trap is escaped because of the further
equal $E$'s or equal $p$'s assumption beyond the first $t \approx x$ approximation.\\
Note that even if the 'equal $v$'s ' assumption is contradictory with production kinematics, 
as shown in \cite{ok} and \cite{GK}, it must embody some approximate truth 
because, for oscillations to be observed, it is important that the two component waves
do not separate too early, and therefore, that they do not have too different velocities. 
$E_1 \approx E_2$ and, say, $m_1 \ll  m_2$ as assumed in \cite{ok} is possible 
only in ultra relativistic situations; but then very small velocity differences make
huge differences in the $\gamma$ factor, so $\frac{\delta v}{v} \ll 1$ is possible, 
though exact equality is excluded; this last impossibility makes the 
loss of coherence and the disappearance of oscillations inescapable after a certain distance.

\section{A 'kinematically correct' assumption}
\subsection{Requirements} A connection between $x$ and $t$ is necessary to 
make predictions about observations which will register a position w.r.t. the creation point, but never 
measure the elapsed time.  This connection should arise from a full quantum treatment
using wave packets or field theory; it would then only be necessary to
propagate the system in time, with the space evolution being taken care of by the wave
equation.\\ 
In the simple plane wave approach, this relation must be imposed because plane waves are of
infinite extent and with two different (mass) states involved one has apparently no
mean of defining a 'center' either by averaging or by stationary phase arguments.\\

The three assumptions detailed above are just different ways of doing this.
No further ingredient is needed in the equal $v$ case, since there is an obvious definition for 
the group velocity. A very likely definition of this velocity has also been used in the equal $p$ 
approximation, thereby avoiding the need of the 'ultra relativistic' $v \approx 1$ and 
$E_1+E_2 \approx 2p$ . Finally, the equal $E$ approximation circumvents the problem by 
yielding a stationary oscillation pattern from the beginning.
\footnote{This makes it the preferred hypothesis of the authors of \cite{Lip1}, \cite{Lip2} because it avoids
ambiguities in transforming theoretical results into observable spatial oscillation patterns.}\\

It is clear now that if we endow the center of the would-be wave 
packet with the average momentum and energy of its components: $\bar{p}=\frac{p_1+p_2}{2}$, 
$\bar{E}=\frac{E_1+E_2}{2}$ and/or postulate that it has
group velocity $v_g = \frac{\bar{p}}{\bar{E}}$, we find back the same expression for the phase 
difference in all three cases; it appears that this postulate embodies all that is needed to yield the usual
 $L_{osc}$:
\beq 
\nonumber 
v = \frac{p_1+p_2}{E_1+E_2}  \Rightarrow & \delta \phi = \delta p x - \delta E t &= (\delta p - \delta E 
\frac{E_1+E_2}{p_1+p_2})x 
=\frac{(\delta p^2 -\delta E^2)x}{p_1+p_2}\\ \nonumber && = \frac{-\delta m^2 x}{2\bar{p}} \eeq

Indeed, this simple hypothesis is a/ implied by the (\ref{eqb}) assumption, b/ identical with the
definition of the group velocity that we have used in the (\ref{eqp}) case and
c/ not required in the (\ref{eqe}) derivation, but not incompatible either.\\

\subsection{The case of $\pi \rightarrow \mu \nu$ at rest}
\label{pimunu}
We show here that our postulate is actually necessary in 
the kinematically most simple case of $\pi \rightarrow \mu \nu$ at rest.\\

From $\; \; \; E_j = \frac{m^2_{\pi}+m^2_j-m^2_{\mu}}{2m_{\pi}}\; $ and $\; E_{\nu}+E_{\mu} 
= m_{\pi}$ one sees that:  
\footnote{Which shows that $\delta E
=\frac{\delta m^2}{2m_{\pi}}$ contrary to the approximation of \cite{dol}}
$$m_1 < m_2 \Rightarrow E_1 < E_2 \Rightarrow E_{\mu,1} > E_{\mu,2}
\Rightarrow p_{\mu,1} > p_{\mu,2}$$ and by momentum conservation $p_1 > p_2$ which, using again
$E_1 < E_2$ shows that $v_1 > v_2$; as expected, all three assumptions are inconsistent with 
kinematics and the third is the worst since $v_1/v_2 = \frac{p_1/p_2}{E_1/E_2}$
 
Now, for oscillations to be
observable, one must have $|p_1 - p_2| < \Delta p$ where $\Delta p$ is the quantal momentum uncertainty due to the 
localization of the source (the $\pi$ in this case), so that one cannot tell which neutrino mass eigenstate 
has  been produced. Consequently, the phase relation is preserved between the
components of the muon state which can be described by a linear superposition of amplitudes. 
 In the plane-wave approximation, this wave function must read something like:
$$\psi =  \cos{\theta} e^{-i(E_{\mu,1}t-p_1x)} + \sin{\theta}e^{-i(E_{\mu,2}t-p_2x)}$$ where $p_1$ and $p_2$ are 
the same as for the two neutrino mass states recoiling against $\mu$. \\
Therefore $|\psi|^2 = 1+\sin{2\theta}\cos(\delta E_{\mu}t - \delta p x)$ and (assuming $\theta \leq \pi/2$)
maximizing the probability gives $x=\frac{\delta E_{\mu}}{\delta p}t=v_{\mu}t$ for the center of 
the (very elementary) wave packet, with $\frac{\delta E_{\mu}}{\delta p}$ 
suggestive of a discrete version of the usual $v_g = \frac{\partial E}{\partial p}$.\\
However in the present case, $\frac{\delta E_{\mu}}{\delta p}=\frac{p_1+p_2}{E_{\mu,1}+E_{\mu,2}}
=\frac{\bar{p}}{\bar{E_{\mu}}}$.\\
But this makes the above postulate for the neutrino velocity more likely, because 
to find this last value for $v_{\mu}$, we have to endow the muon state  
with the mean momentum $\bar{p}$ and energy $\bar{E_{\mu}}$ of its two
components, and likewise for the neutrino in order to enforce energy and momentum conservation in
the decay; in turn, this implies our neutrino velocity. Of course, other ways of 
averaging could do 
the job as far as conservation laws are concerned; but it is hard to see how  $\theta$-dependent
averages of $E$ and $p$ could yield back the same ($\theta$-independent) expression for the velocity 
of the muon obtained by the requirement of maximum probability.
\footnote{Which is but a simplified and discretized version of standard stationary phase arguments  
used to define the center of a wave packet}\\
Certainly, assuming $\Delta p > |p_1-p_2|$ and using a superposition of two plane waves 
with $p=p_1$ and $p=p_2$ for the muon is not perfectly consistent, but not
worse either than what is done on the neutrino side. The reason for which we had to use the muon argument is
evidently that the ill-defined oscillating state (\ref{act}) has a constant modulus. Is is not possible
to make any direct statement about its location in space-time without first defining its flavour content
(as is done to calculate the oscillation length).

\section{Restrictions and conclusion}
\label{lesson}
The shortcomings of the plane wave description
of neutrino oscillations were discussed long ago \cite{R} ;
the necessity of using wave packets (see e.g.\cite{KRG})
or field theory (\cite{GMS},\cite{Car1}) and of incorporating 
the neutrino production and/or detection processes in the description of the phenomenon (\cite{R},\cite{C}) 
has been the subject of many works and still feeds a continuing debate.
However, the results of these more elaborate treatments always imply,
up to damping factors, the oscillation pattern described by (\ref{exact})
in the relativistic limit, whenever oscillations are not washed out by resolution or loss of coherence
 \footnote{Of course we discount incorrect treatments of the equal velocity assumption}.

The above analysis explains the robustness of this classical formula by exposing 
the only ingredient necessary to retrieve it; this turns out to be more general than  
what is used in standard derivations and does not contradict production kinematics.\\
As a by-product, it also shows that the variable to be used in the oscillation length formula
is the momentum rather than the energy, should the distinction apply. However this would happen in
the non-relativistic regime where, due to phase-space limitations and admixture of 'wrong' helicity
states, an oscillation probability disconnected from the rest of the process has a very limited 
meaning \cite{Car}, see also \cite{GKM}.

Obviously, the restriction to two flavours is a weakness of the analysis presented here; 
it must be remarked, however, that the oscillation length has a clear
physical interpretation only in this case. The corresponding parameters become unavoidably 
entangled with mixing matrix elements when dealing with more than two flavours.
Nevertheless, it is our hope that the simplicity of the arguments and results 
presented here may shed some light on the general case.

\newpage
\end{document}